\documentclass[sigconf,screen,nonacm] {acmart}

\AtBeginDocument{%
  }

\setcopyright{cc}
\setcctype[4.0]{by}

\begin{document}

\title{Social Facilitation of Creative Reflection: AI-agents and Humans}

\author{Olga Sutskova}
\authornote{Equal Contribution}
\orcid{0000-0003-3222-524X}
\affiliation{%
  \institution{University of the Arts London}
  \city{London}
  \country{UK}
}
\email{olga.sutskova@arts.ac.uk}

\author{Corey Ford}
\authornotemark[1]
\orcid{0000-0002-6895-2441}
\affiliation{%
  \institution{University of the Arts London}
  \city{London}
  \country{UK}
}
\email{c.ford@arts.ac.uk}

\begin{abstract}
Social collaboration can support people's reflection and is a crucial component of creativity. Creative technologies have been designed to support more collaborative ways of working, including using AI to simulate social partners. As human-AI creative collaborations increase, further investigation is needed into how different social interactions influence creative reflection and at which stage a social intervention is crucial to improve creative outcomes. Considering that non-verbal communication is the bedrock of human cognition and influence, non-verbal social dynamics should be examined in detail in the age of AI-companionship. For example, during social interaction, the social facilitation effect describes how the mere presence or observation of others influences how a person behaves and feels in the context. Whether changes in technology-mediated social environments influence how people reflect on their creative work needs further exploration, as does whether social AI-companionship elicits similar effects as humans. This paper discusses how theoretical mechanisms relating to social facilitation could influence creative practice and reflection, proposing ways of further testing these effects. 
\end{abstract}




\maketitle

\textbf{Reference:}\newline
Olga Sutskova and Corey Ford. 2026. Social Facilitation of Creative Reflection: AI-agents and Humans. In \textit{Proceedings of The First Reflection in Creative Experience (RiCE) Workshop (RiCE W1)}. ACM Creativity \& Cognition 2026, London, UK.

\section{Introduction}

Models of reflection often focus on individual mental processes, including moments of resolving difficulty \cite{dewey_how_1933,boyd_reflective_1983,atkins_reflection_1993} and transformative change \cite{atkins_reflection_1993,boud_promoting_1985}. In creative contexts, reflection has been characterised as occurring both during action \cite{schon_reflective_1983,candy_creative_2019} and on process \cite{Ford2023}, varying across stages of creative work (e.g. \citet{Ford2024}). While these models emphasise individual cognition, reflection has also been understood as inherently social \cite{bentvelzen_revisiting_2022}. There are several examples of how reflection occurs in dialogue between people in creative contexts, e.g., during design criticism \cite{Bardzell2013}. Central to social scaffolding of reflection is the notion that different backgrounds, personalities and preferences can expose people to different perspectives, allowing discussion on blind spots which are otherwise not exposed through solitary work \cite{Tsang2007}.

Creative technology has also been used to facilitate reflection by encouraging social collaboration. For example, Mosaic \cite{kim_mosaic_2017} is a Creativity Support Tool (CST) where communities of creative practitioners are able to comment on documentation of their creative process. With recent advances in AI renewing discussions on the use of computers in creativity, there are also several examples of tools which describe AI as a possible creative collaborator. 
Although, in this paper, we adopt the view of AI more as a social tool rather than an equal collaborator e.g. see cf. Shneiderman's Human-Centred AI \cite{shneiderman_human-centered_2022}. This is due to current AI systems exhibiting several limitations with regard to creating a truly equal collaborative experience. For example, AI chatbots typically lack the agency and genuine interest in a creative artefact integral to creative co-creation \cite{Yin2026}. AI also often displays sycophantic behaviour, lacks constructive critique \cite{Li2026}, and often encourages early convergence (limited optimal solution) rather than divergent (diverse creative) thinking  \cite{Lu2022}. Nonetheless, as a social tool, an AI-companion can navigate the creative through a structured, reflexive framework. An AI-companion could also simulate creative critique in the language patterns of publicly known critics, if trained to do so. Whether this type of interaction is meaningful enough to spark a creative shift and in-depth reflection, similar to another person, needs further investigation.

\section{The Need for Psychological Theories in HCI}

Other people's input, either in person or mediated through technology, can be central to meaningful reflection. Outside of creative HCI research, the Psychological (Cognitive) Sciences have described different and overlapping mechanisms of human cognitive processes which relate to reflection. Interactive social reflection practices involving discourse in particular are linked to both metacognitive processes and theory of mind practices (thinking about what others are thinking), incorporating others' perspectives within one's own knowledge frameworks \cite{frith_2012_mechanisms}. More implicit social contexts, such as co-presence of others, have been shown to orient participants' attentional resources towards the self, adjusting behaviours accordingly and changing trajectories of their performance ability \cite{Guerin1983}. Such theories are underexplored in creative HCI, yet could complement theories used to examine creative practice, such as group flow \cite{sawyer2014group} or mutual engagement \cite{bryan-kinns_identifying_2012}.

This discussion is timely, as the calls for more psychological research contributions to human-technology interaction and specifically human-AI interaction grow \cite{xu_toward_2019}. As researchers with a background in Social Cognition and Technology (Cognitive Psychology) and Creative Reflections (CSTs), we propose an exploration of how some of the canonical psychological effects usually observed in human-to-human social contexts can be further explored in the age of social AI. 
This paper thus outlines some research directions for examining social cognition and affect that could influence reflection from a Cognitive Science perspective. The aim is to provoke interdisciplinary discussion and collaboration on the topic of technology-mediated and AI-driven social facilitation and its impact on creative practice and reflection. We introduce theories on the social cognition process and their relationship to reflection below. 

\section{Social Cognition Processes}

Social cognition processes can be discourse-based (as often used in LLM analyses), but can also focus on more implicit non-verbal aspects of human-human communication \cite{healey2021human}. In this report, we are interested in implicit non-verbal social influences, in particular, the \emph{Social Facilitation Effect (SFE)}. SFE is a social phenomenon where people's behaviour, performance, and state of awareness change in social contexts compared with their lone working \cite{Allport1920, bond1984social}. The canonical effect usually focuses on improved performance on easy tasks (facilitation), and detriment in performance on difficult (inhibition) tasks in social contact versus alone. The effects can be elicited through two distinct mechanisms \cite{Guerin1983}, either by someone observing the performer (the Audience Effect) or when someone is co-present with the same performance space (the Mere Presence Effect). Whether reflection ability could be affected in a similar pattern is yet to be tested. 

Additionally, the co-presence of other people is shown to affect people's perception of events. For example, watching films or listening to music together amplifies the experience \cite{Boothby2014}, whilst being watched skews the perception of time and evaluation of your own work \cite{Steinmetz2025, Steinmetz2016BeingObserved}. This relates to the theory of group flow \cite{sawyer2014group}, which is often leveraged in creative HCI research, describing a state of optimal performance where teams become immersed in a shared activity and time fades away. Neuroscientific evidence from fNIRS studies finds that co-present activities, such as watching films together, lead to synchronisation of brain signals even when people are not directly interacting with each other \cite{DeFelice2024}. This could suggest that the presence of others sets people in a certain state of social equilibrium. We speculate that this equilibrium could be the driving force behind creative flow during co-creation. Whether there is a certain level of task difficulty at which creative practice and reflection benefit or deteriorate, and whether technology could moderate or mediate these effects, should be investigated further. The findings could enhance socially mediated creative practice alongside humans and social-AI tools. 

\subsection{SFE and Creative Reflection with AI}

Considering that most work on SFE's has been done within human-to-human scenarios, there is an opportunity to inspect how SFE (audience effect and mere presence effect) mechanisms affect creativity and reflective cognition, also alongside AI companions. For example, Sutskova et al. \cite{Sutskova2022} found that videoconference co-presence alongside an embodied AI-agent and a human-companion both elicited improvement in performance accuracy on newly learned cognitive tasks. When performance was monitored, however, only real-human observation elicited a performance change. This could suggest that there is a cut-off point at which an AI-companion might not suffice, and shows that human presence is essential to elicit behavioural change. Older frameworks exploring interaction with AI-agents versus human virtual companions note that agents might never have the same impact as humans, due to their lack of real-world influence over people's lives \cite{Blascovich2002}. There are some examples of how different creative activities are influenced by the SFE effect, including changing learners anxiety when practising musical instruments \cite{song2020robot}, changing performers body expressions in dance \cite{shikanai2014dancers}, and changing the number of ideas generated with chatbots \cite{Hwang2021Ideabot}. However, these examples do not compare the impact on creative reflection and there is scope for a wider exploration of different SFE mechanisms. 

With LLM use becoming more pervasive in everyday life and improvements in AI-embodiment and independent agentic behaviours, the trends of how agenting influence might be perceived and impact human lives could be changing. For example, people are now perceiving LLM-agents as more independent, and might be susceptible to their influence much more than when agents served mainly limited functions such as gaming NPCs or basic customer service bots. The LLM agents are nowadays used as trustworthy sources on personal and relationship advice \cite{Tseng2026}. Therefore, there needs to be an exploration of drifting perceptions of AI influence. This could be done by applying and expanding the psychological theories and frameworks towards these augmented social connections. For example, by testing whether more "intelligent" or embodied LLM's can elicit stronger AI-co-presence (mere presence effect) and the sense of being evaluated by these systems (audience effect).

Applying classic theoretical frameworks, it is possible to test when people perceive AI-companions similar to humans, and when that perception breaks down. And then, whether reflection can be deepened with the co-presence of others (human or AI-agents), or whether it is only amplified through the anticipation of the evaluation from other people (audience effect). Importantly, the investigation needs to account for which stages of reflection and creative processes are facilitated by these mechanisms (e.g. during convergent or divergent thinking stages \cite{designcouncil2005doublediamond}) and at which point social context is detrimental, and solitude is preferred.  

The SFE is also susceptible to effects beyond just objective performance ability when alone versus when in the presence of others. For example, the effects are heavily susceptible to subjective beliefs about self, own abilities and personality, but also beliefs about the expertise and intentions of others in the context \cite{uziel2007individual}. The effect is also warped by whether the co-presence or monitoring is perceived as positive or negative \cite{seta1995when}. The positive or negative perception of social others influencing a persons physiological states (\cite{blascovich1999social}).  

Similarly to SFE, creative practice and reflection are also heavily reliant on processes such as beliefs about self and own efficacy, others competencies and judgement, the nature of their evaluation in the context. Therefore investigating how social presence might influence these processes and how AI tools might mitigate the negative effect of such processes could a be an interesting frontier for psychological and HCI research collaboration.  

Stepping away from the human-AI impact discourse, the question of what co-presence entails is still open. With researchers proposing co-presence being anything from an interactive interface involving another person or agent remotely, to haptic presence, to audio and visual presence, to the idea that presence can only be evoked through physically sharing an environment \cite{Lee2004}. Collaborative and creative user interfaces could be one important way to experimentally test these levels of impact, systematically separating different social modalities \cite{SutskovaOlga2026DMPa, Sutskova2022, SutskovaOlga2023CIoS}.

\section{Mechanisms to Elicit Reflection}

To be able to examine effects such as SFE and co-presence on reflection, we need to understand the mechanisms which elicit reflection in people. These mechanisms can be internal (relating to cognitive processes in the brain) or external (sparked by the environment or technology). 

Some mechanisms which trigger reflection are subtle and relate to more in-the-moment creative activity. For instance, people may reflect-in-action \cite{schon_reflective_1983} and conduct small experiments based on their making practice. Whilst people may not reflect consciously \textit{per se}, their decision-making and actions relate to their tacit knowledge of their making processes, built up over time. However, reflection has also been described as triggered in moments of breakdown \cite{baumer_reflective_2015} (when an interaction is interrupted and a persons awareness of their actions becomes apparent), or during a moment of difficulty \cite{dewey_how_1933} or inner discomfort \cite{boyd_reflective_1983}. This breakdown can be elicited through purposefully adding ambiguity into designs \cite{gaver_ambiguity_2003} or by presenting information as untrustworthy \cite{norman_things_1993}. Reflection can also be elicited through encouraging comparison, either to other people \cite{bentvelzen_revisiting_2022} or between people's current and previous experience \cite{boud_promoting_1985, schon_reflective_1983}. New experiences can also trigger reflection, where the novelty encourages people to reflect from a fresh perspective \cite{kolb_experiential_1984}. Given this, we suggest that social mechanisms could introduce a type of disruption to trigger a cognitive challenge either through the environment or in a person's own actions and perceptions of such. Indeed, disruptions could be stimulated through the introduction of social agency of others, and there is an opportunity to leverage mechanisms such as SFE to this effect.




\subsection{SFE Mechanisms for Reflection}

Arguably, SFE is an effect of social disruption, as it introduces another social agent into an otherwise solitary activity of creative or reflective process. Through mechanisms such as the Mere Presence effect and the Audience Effect, individuals' attention is no longer focused on just the process, but on the environment around them and who they might be sharing this environment with. In psychological sciences, this state of social processing is attributed to second-person cognition \cite{Schilbach2013}. 

The Audience Effect, as an SFE mechanism dependent on being observed by others, drifts the attention towards the anticipation of others' evaluation of the protagonist's performance of self and their abilities \cite{Hamilton2016, GuerinInnes1982}. These social processes interact with beliefs about self-efficacy \cite{Bond1982}, the efficacy of the audience \cite{Geen1991} and are believed to elicit certain levels of reputation management when watched \cite{Wolf2015, Cottrell1968}. This level of disruption could, in principle, trigger the reflection mechanisms relating to inner discomfort or create a certain level of breakdown that forces the individual to step out of their comfort zone. The beliefs about what the observer might be thinking about the creative or reflective process as they observe could, on their own, elicit new ideas and deeper insight into one's own work. Notably, as per the canonical SFE effect, task difficulty is an important aspect of how performance is affected. Therefore, the observation might only benefit these processes when people are under a lower cognitive load stage.

The Mere Presence Effect is an SFE mechanism dependent on perceiving someone else's co-presence during some level of activity. The presence does not need to be collaborative or competitive. Researchers propose that this co-presence can be established through different means, from physically sharing a room to more singular perceptual mechanisms, such as auditory, haptic, interactive display, etc., \cite{Lee2004}. The psychological processes evoked by co-presence are often related to the state of preparedness, vigilance and social alertness \cite{Guerin1983}. These states are believed to keep a person in a higher state of psychological arousal, eliciting divided attention \cite{Sanders1981}. Similar to the Audience Effect, social processes elicit a certain attentional shift to the self. However, unlike the Audience Effect, the Mere Presence Effect is not reliant on others' judgment, more so on the monitoring of self and shared environment and related heightened alertness \cite{Guerin1983, Guerin1986}. Slightly elevated levels of such arousal are believed to lead to more concentrated attention, leading to better focus. Too high levels could lead to scattered attention and less focus \cite{Eysenck2007}. Therefore, similar to AE, simulating an optimal level of co-presence could potentially lead to a more focused creative and reflection process. This notion of focus connects with theories of optimal states for creative practice cf. group flow theory \cite{sawyer2014group}.

There should be further investigation on whether AI tools could be used to understand individuals' levels of cognitive load during creative and reflection processes, and then potentially propose a social facilitation intervention or a social rest. More research should aim to understand which social interventions would be most beneficial to spark additional creative and reflective outcomes, and at which levels of these processes. Additionally, psychological testing is required to understand whether AI-social companions could be in any way useful in these practices, with human well-being in mind.


    
    
    
    



\section{Closing Remarks}

In this article, we have proposed that there is a need for more unified theories of socially mediated reflection and creative practices. This is especially important now when social communication is augmented through social AI. To achieve this goal, we call for stronger collaborations between psychological sciences and HCI. 
In this brief report, we have summarised how our fields of creativity-related HCI and psychology (cognitive science) can work together to bridge the theoretical gaps. We have briefly demonstrated the theories and mechanisms related to social facilitation and creative reflection in both fields, and how these theories should be expanded towards AI tools and companionship. In particular, how social mechanisms could be leveraged to trigger mechanisms for reflection such as breakdowns. 
We would like to expand on these discussions further alongside the multidisciplinary attendees of the ACM Creativity and Cognition conference (2026), as part of the RiCE workshop. The proceedings and discussion in this workshop will be used to develop future collaborations between cognitive science and CST research in the field of augmented social cognition and creative practice and reflection.

\bibliographystyle{ACM-Reference-Format}
\bibliography{references}

\appendix

\end{document}